\documentclass[paper]{JFM-FLM_Au}

\providecommand{\dd}{\mathrm{d}}
\makeatletter
\@ifundefined{theorem}{\newtheorem{theorem}{Theorem}}{}
\makeatother
\newtheorem{lemmax}{Lemma}

\makeatletter
\def\ps@titlepage{\leftskip\z@\let\@mkboth\@gobbletwo\vfuzz=5\p@
  \def\@oddhead{\vbox{\vspace*{-4pt}\hbox to \textwidth{\@j@urnal \hfill}}}
  \def\@evenhead{\vbox{\vspace*{-4pt}\hbox to \textwidth{\@j@urnal \hfill}}}
  \def\@oddfoot{\hbox to \textwidth{\hfill{\cppagefont\ifx\@volume\undefined\else\textbf{\@volume}\fi\ \ifx\@issue\undefined\ifpaper A\else X\fi1-\else\ifpaper A\else X\fi\@issue-\fi\thepage}}}
  \def\@evenfoot{\hbox to \textwidth{{\cppagefont\ifx\@volume\undefined\else\textbf{\@volume}\fi\ \ifx\@issue\undefined\ifpaper A\else X\fi1-\else\ifpaper A\else X\fi\@issue-\fi\thepage}\hfill}}
  \def\sectionmark##1{}%
  \def\subsectionmark##1{}%
}
\def\pagelimitfooter{\hbox to \textwidth{{\cppagefont\ifx\@volume\undefined\else\textbf{\@volume}\fi\ \ifx\@issue\undefined\ifpaper A\else\ifrapid R\else\ifpersp P\else\iffof F\else X\fi\fi\fi\fi1-\else\ifpaper A\else\ifrapid R\else\ifpersp P\else\iffof F\else X\fi\fi\fi\fi\@issue-\fi\thepage}\hfill}}
\makeatother

\lefttitle{Y. S. Park}
\righttitle{Journal of Fluid Mechanics}

\title{Exact linearisation of the nonlinear shallow-water equations
over a moving bottom: classification, wave generation and run-up}

\author{Yong Sung Park\aff{1,2}}

\affiliation{\aff{1}Department of Civil, Urban and Environmental Engineering,
Seoul National University, Seoul 08826, Republic of Korea
\aff{2}Institute of Construction and Environmental Engineering,
Seoul National University, Seoul 08826, Republic of Korea
}

\corresau{Yong Sung Park, \email{dryspark@snu.ac.kr}}

\begin{document}

\maketitle

\begin{abstract}
The Carrier--Greenspan hodograph transformation linearises the nonlinear
shallow-water equations on a beach of constant slope. We extend the transformation to a moving
bottom and identify the family of bottom motions that preserves the
Carrier--Greenspan invariants. The bottom gradient must be uniform in
space, that is, $h(x,t)=\rho(t)x+\mathcal{B}(t)$, and constant $\rho$ returns the classical
slope. For this family a horizontally accelerating frame removes the forcing. The
bottom motion then disappears from the hodograph equation, and two quadratures
recover the physical variables. If the gradient varies in time, the bed displacement grows
without bound offshore and the far field cannot remain at rest. The domain must
therefore be finite, although the pivot may be placed anywhere. A bottom-tilting wave maker realises this family. Closed forms follow for the
hinge response, for the crest-to-trough steepness of the generated wave, and for
the leading nonlinear correction.
The correction depends on the point to which the steepness is referred, and so
differs between a leading-elevation and a leading-depression wave. Because the
hinge is at the beach toe, these closed forms provide the incident wave required
by the classical plane-beach solution. The run-up can therefore be predicted from
the plate motion alone. The
steepness is predicted without a fitted constant and agrees with the reference computations. Sixty-one runs over a level bed,
reported here for the first time, carry the test to deeper water and longer plate
motions. The measured run-up follows
the predicted run-up, with an offset that the inviscid theory does not account for.
\end{abstract}


\section{Introduction}\label{sec:intro}

The nonlinear shallow-water equations with a moving shoreline have few exact
solutions. \citet{CarrierGreenspan1958} solved one-dimensional motion on a beach
of constant slope. Their hodograph transformation turns the two nonlinear
equations into one linear equation, and the moving shoreline into a coordinate
line. \citet{Thacker1981} solved two-dimensional motion in a basin of parabolic
profile.

Later work extended the Carrier--Greenspan solution to other
geometries. Channels of arbitrary cross-section, $U$-shaped bays and general
initial--boundary-value problems have all
been solved \citep{RybkinPelinovskyDidenkulova2014, Rybkin2021}.
\citet{Camassa2022} followed the singularities of the free surface over a bottom
that varies in space, and noted that the transformation linearises the
equations when $h_{xx}=0$, which is the condition of \S\,\ref{sec:class} for a
bottom at rest.
\citet{AntuonoHoggBrocchini2009} studied a bottom tilted impulsively and then
held fixed. \citet{AntuonoBrocchiniSAM2010} treated a fixed non-planar bottom by
perturbation, and \citet{AntuonoBrocchini2010} gave an alternative to the
hodograph route in physical variables. In each of these the bottom is at rest
while the wave on the free surface propagates.

Previous studies of a continuously moving bottom have followed two approaches,
linear theory and approximation within the nonlinear equations.
\citet{TuckHwang1972} solved the linear shallow-water equation on
a uniform slope with a prescribed bottom displacement by Laplace and Hankel
transforms. \citet{LiuLynettSynolakis2003} extended the solution and marked its limit. They expect nonlinearity to
dominate once $\tan\beta/\mu$ falls below unity, with $\beta$ the beach angle
and $\mu$ the aspect ratio of their scaling. \citet{DutykhDias2007} worked in linearised potential flow and argued
that the time scale of the bottom deformation matters, citing slowly spreading
uplift and the slow faulting motion in the northern part of the 2004
Sumatra--Andaman source. They wrote the source as a static sea-floor deformation multiplied by one of four time functions, the last of which is a linear ramp.
In the second approach,
\citet{OzerenPostacioglu2012} applied the Carrier--Greenspan transformation to a
moving bottom and noted that the transformed equation is no longer strictly
linear once the depth depends on position as well as time.
\citet{MadsenHansen2012} kept the nonlinear equations for a bottom obstacle
translating over constant depth and reduced the two families of characteristics
to one. They performed the elimination analytically and reported the result to be
almost as accurate as the two-family system.

In this paper, we identify all bottoms $h(x,t)$
for which the Carrier--Greenspan transformation remains exact. The only
admissible family is
$h=\rho(t)x+\mathcal{B}(t)$. The bottom-tilting wave maker of \citet{LuParkCho2017b}, a
flat plate hinged at one end and driven over its whole length, realises the
formulation, and \S\,\ref{sec:comp} compares the theory with measurements made
by \citet{LuParkCho2020}.

The hinge of the wave maker stands at the toe of a plane beach. The exact
solution provides the incident wave at the beach toe, where the classical run-up
problem begins. Following \citet{Synolakis1987}, we can then predict the run-up
without measuring or numerically computing the wave at the toe.

Two further comparisons are worth noting. \citet{TintiTonini2005} used the
Carrier--Greenspan transformation for tsunamis raised by nearshore earthquakes
over an ocean of constant bottom slope. The earthquake appears as an initial condition. The co-seismic sea-floor
displacement is transferred to the initial free surface, and the bottom is fixed
henceforth. In the present paper, the bottom is still moving while the wave is
in motion. Run-up has been examined from other directions,
among them the normal-mode analysis of \citet{Postacioglu2017} and the
initial-value treatment of \citet{CarrierWuYeh2003}.
\citet{MaranzoniMignosa2019} obtained exact solutions of the shallow-water
equations under a horizontally uniform body force, in the Thacker lineage, with
the bottom at rest. One result below (\S\,\ref{sec:reduce}) is that a tilting
bottom and a horizontally uniform body force drive the same flow.

The paper is organised as follows. \S\,\ref{sec:class} states the classification
theorem. \S\,\ref{sec:reduce} derives the reduction and gives the general
solution in quadratures. \S\,\ref{sec:finite} explains why the domain must
be finite and why the pivot may be placed anywhere along the bottom. \S\,\ref{sec:runup} treats
run-up. \S\,\ref{sec:gen} turns to wave generation, where the theory predicts a
measured quantity without an adjustable constant. \S\,\ref{sec:comp} makes the
comparison with experiment and predicts the run-up from the plate motion. \S\,\ref{sec:concl} summarises the results and
suggests further research directions.

\section{Formulation and classification}\label{sec:class}

Consider a long wave over a bottom which is free to move while the wave propagates.
The origin is at the initial shoreline and $x$ increases offshore. Let
$d(x,t)$ be the total depth, $u(x,t)$ the depth-averaged velocity and
$z=-h(x,t)$ the bottom, so that the free surface stands at
$\zeta=d-h$. The nonlinear shallow-water equations are
\begin{equation}
d_t+(du)_x=0,\qquad u_t+uu_x+g\,d_x=g\,h_x .
\label{eq:nsw}
\end{equation}
Note that a subscript $t$, $x$, $X$, $\lambda$, $\sigma$ or $r$ denotes partial
differentiation henceforth, and every other subscript is a label.
With $c=\sqrt{gd}$ the characteristic form is
\begin{equation}
\bigl[\partial_t+(u\pm c)\partial_x\bigr](u\pm2c)=g\,h_x(x,t).
\label{eq:char}
\end{equation}
The upper and lower signs belong to the characteristic families $C_+$ and
$C_-$. Only $g\,h_x$, the component of gravity along the bed, brings the bottom into
the equations. The transformation of \citet{CarrierGreenspan1958} rests on the
two combinations
$u\pm2c$ staying constant along their characteristics. The constancy is maintained
as long as $h_x$ is independent of $x$.

\begin{theorem}[Classification]\label{thm:class}
In the wetted region $c>0$, let $P_\pm(u,c,x,t)$ be functions which stay constant
along the characteristics of their own family, for every solution of
\eqref{eq:nsw}, and let $\partial P_\pm/\partial(u\pm2c)\neq0$. Such a pair
exists if and only if
\begin{equation}
h_x(x,t)=\rho(t)\qquad\Longleftrightarrow\qquad h(x,t)=\rho(t)\,x+\mathcal{B}(t),
\label{eq:family}
\end{equation}
and the invariants are then
\begin{equation}
u\pm2c-V(t),\qquad V(t)=g\!\int^{t}\!\rho(t')\,\dd t' .
\label{eq:invariants}
\end{equation}
\end{theorem}

\begin{figure}
  \centering
  \includegraphics[width=\textwidth]{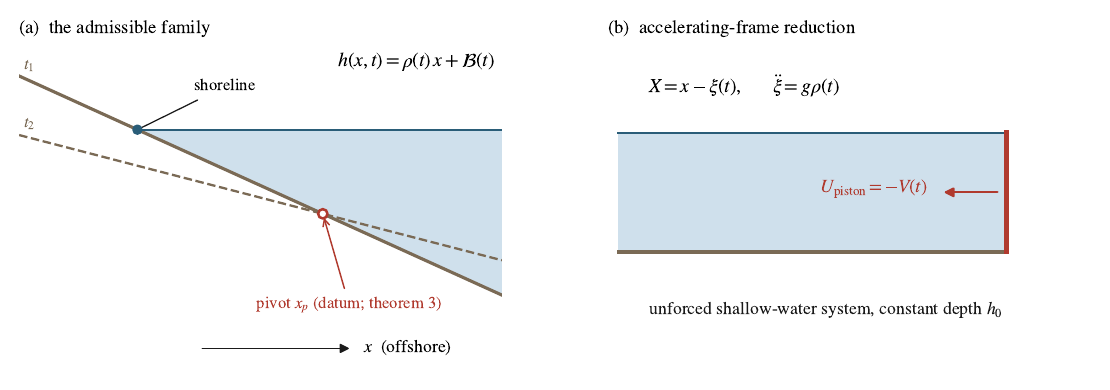}
  \caption{(\textit{a}) The admissible family $h=\rho(t)x+\mathcal{B}(t)$ at two instants;
  the two bottoms cross at the pivot, which sets the datum
  (theorem~\ref{thm:pivot}). (\textit{b}) The accelerating-frame reduction:
  constant depth, unforced equations, and a piston at the wall moving with
  velocity $-V(t)$.}
  \label{fig:geometry}
\end{figure}

\noindent Sufficiency follows at once from \eqref{eq:char}. Necessity can be shown in
two steps. First, \eqref{eq:char} fixes the variation of $u+2c$ along $C_+$ but
leaves that of $u-2c$ unrestricted. The invariant must therefore depend only on
$u+2c$. Requiring this reduced function to remain constant then separates the
$x$-dependence (Appendix~\ref{app:hodo}).
Setting $\rho=\mathrm{const}$ recovers the beach of fixed slope. Every other bottom in the family tilts as a whole
about some pivot, as in figure~\ref{fig:geometry}(\textit{a}). A
transformation built on other dependent variables would carry its own
admissible set.

\section{Reduction and general solution}\label{sec:reduce}

\subsection{The accelerating frame}

For constant $\rho$, both the family and the following reduction are known.
\citet{AntuonoHoggBrocchini2009} solved the fixed-slope case, $\dot\rho=0$, in a
finite domain initially at rest. They showed that the solution eventually becomes
multivalued and must then include discontinuities. \S\,\ref{sec:class} adds that the reduction survives a
continuously varying $\rho$, and \S\,\ref{sec:gen} works in the window before
those discontinuities appear.

A bottom in the family \eqref{eq:family} has $h_x=\rho(t)$ at every station, and
at any instant the forcing accelerates the whole water column equally. Take a frame that moves with the acceleration:
\begin{equation}
X=x-\xi(t),\qquad U=u-V(t),\qquad \dot\xi=V,\qquad \ddot\xi=g\rho(t).
\label{eq:frame}
\end{equation}
Since $\partial_t|_x=\partial_t|_X-\dot\xi\,\partial_X$ and $u=U+V$, the two
equations of \eqref{eq:nsw} become
\begin{equation}
d_t+(dU)_X=0,\qquad
U_t+UU_X+g\,d_X=g\rho(t)-\ddot\xi=0 .
\label{eq:unforced}
\end{equation}
The apparent force of the frame cancels the bottom forcing exactly, so in the
$(X,t)$ frame the shallow-water equations are \emph{unforced}
(figure~\ref{fig:geometry}\textit{b}). A bottom that tilts and a horizontally uniform
body force drive the same flow.

\subsection{Hodograph equation}

With $\lambda=2U$ and $\sigma=4c$ the invariants \eqref{eq:invariants} are
constant along their characteristics, and
\begin{equation}
X_\lambda=U\,t_\lambda-c\,t_\sigma,\qquad
X_\sigma=U\,t_\sigma-c\,t_\lambda .
\label{eq:Xsys}
\end{equation}
Cross-differentiation yields
\begin{equation}
\sigma\bigl(t_{\lambda\lambda}-t_{\sigma\sigma}\bigr)-3\,t_\sigma=0 ,
\label{eq:EPD}
\end{equation}
which is the Euler--Poisson--Darboux equation. Equation
\eqref{eq:EPD} is independent of $\rho(t)$, so one linear equation serves the
whole family (see Appendix~\ref{app:hodo}).

\subsection{Recovery of the physical plane}

With $X=\tfrac12\lambda t-W$, the system \eqref{eq:Xsys} integrates to
\begin{equation}
W_\sigma=\frac{\sigma}{4}t_\lambda,\qquad
W_\lambda=\frac{t}{2}+\frac{\sigma}{4}t_\sigma ,
\label{eq:W}
\end{equation}
whose integrability condition is exactly \eqref{eq:EPD}. The physical variables
follow as
\begin{equation}
x=\xi(t)+X,\quad u=\frac{\lambda}{2}+V(t),\quad
d=\frac{\sigma^{2}}{16g},\quad
\zeta=\frac{\sigma^{2}}{16g}-\rho(t)\,x-\mathcal{B}(t),
\label{eq:recon}
\end{equation}
with the shoreline at $\sigma=0$. Equations
\eqref{eq:EPD}--\eqref{eq:recon} are the general solution for the family.

\section{The finite domain}\label{sec:finite}

\subsection{Incompatibility with an infinite domain}

A bottom whose uniform gradient changes in time moves through a vertical
distance proportional to $x$, and far offshore the displacement grows
without bound. Only a domain of finite length keeps the displacement bounded.

\begin{theorem}[Far-field incompatibility]\label{thm:farfield}
Let $h=\rho(t)x+\mathcal{B}(t)$ with $\rho>0$ on $x>x_1$, and let the free surface and
its rate of change stay bounded far offshore. If $\dot\rho\neq0$ then $u$ cannot tend to zero as
$x\to\infty$; instead $u\sim-\dot\rho\,x/(2\rho)$.
\end{theorem}

\noindent Integrating the continuity equation of \eqref{eq:nsw} over $(x_1,x)$ leads to
\begin{equation}
\frac{\dd}{\dd t}\!\int_{x_1}^{x}\!d\,\dd x'+\bigl[d\,u\bigr]_{x_1}^{x}=0 ,
\label{eq:volume}
\end{equation}
and with $d=\zeta+\rho x+\mathcal{B}$, and with $\zeta$ and $\zeta_t$ bounded, the integral
changes at a rate that grows as $\tfrac12\dot\rho\,x^{2}$. The tilting bed
displaces volume at a rate proportional to $x^{2}$. This volume must pass through
a cross-sectional area $d\sim\rho x$, giving
$u\sim-\dot\rho x/(2\rho)$. The velocity is therefore unbounded unless $\dot\rho=0$.

Linear theory has the same obstruction. For a change in still depth of our
family, $h_b=\rho_b(t)\,x$, the linear forced equation of
\citet{TuckHwang1972} and \citet{LiuLynettSynolakis2003} reads
\begin{equation}
\zeta_{tt}-g\tan\beta\,\bigl(x\zeta_x\bigr)_x=-\partial_{tt}h_b ,
\label{eq:lls}
\end{equation}
in which $h_b(x,t)$ is the prescribed change in still depth, $\rho_b$ its
gradient, and $\tan\beta$ the beach slope. The forcing then grows
linearly in $x$, and the Hankel transform on which their solution rests no longer converges.

\subsection{Pivot invariance}

\begin{theorem}[Pivot invariance]\label{thm:pivot}
For the same initial depth, the flow $(u,d)$ is independent of the choice of
pivot $x_p$. The bottom enters \eqref{eq:nsw} through $h_x$ alone, so $\mathcal{B}(t)$
plays no role in the dynamics.
\end{theorem}

\noindent The pivot only sets the datum. Moving it changes the description of
the bottom but not the flow. The free surface does change,
since $\zeta=d-h$. Placing the pivot at the toe of a laboratory wave maker, for example, keeps the bottom displacement
bounded over the whole wetted region.

The linear problem \eqref{eq:lls} behaves in the same way. For the same family it
admits the exact particular solution $\zeta_{\rm par}=-\rho_b(t)x+\Xi(t)$ with
$\ddot\Xi=-g\tan\beta\,\rho_b$, which carries the tilt with the sign it has in
\eqref{eq:recon}, and the remainder is the classical unforced
problem. The offset $\Xi$ is the linear counterpart of the frame displacement
$\xi$, and its two constants of integration play the part of the pivot datum.

\subsection{Boundaries in the hodograph plane}

The end wall of the tank is impermeable, so $u=0$ there. The condition maps to
\begin{equation}
\lambda=-2V(t),
\label{eq:wall}
\end{equation}
and when $V$ is monotone the time along the wall is an explicit function of
$\lambda$. After the plate returns to the horizontal, $V$ remains constant and
the wall maps to a coordinate line. The boundary is curved only while the wave is
being generated.

\section{Run-up}\label{sec:runup}

The depth vanishes at the waterline, which puts the shoreline at $\sigma=0$.
Shoreline motion is fixed by the behaviour of $t$ near the line.
Regular solutions of \eqref{eq:EPD} expand in even powers of $\sigma$. Writing
$t=\sum_n a_n(\lambda)\sigma^{2n}$, \eqref{eq:EPD} holds if and only if
$a_{n+1}=a_n''/[4(n+1)(n+2)]$, so that with $a_0\equiv T$
\begin{equation}
t(\lambda,\sigma)=T(\lambda)+\frac{T''(\lambda)}{8}\sigma^{2}
+\frac{T''''(\lambda)}{192}\sigma^{4}+O(\sigma^{6}).
\label{eq:texp}
\end{equation}
The single free function $T(\lambda)$ is the time at the shoreline.
Substituting \eqref{eq:texp} in \eqref{eq:W} and evaluating at $\sigma=0$ produces
the time, position and velocity of the shoreline, all parametrised by $\lambda$,
\begin{equation}
t_{\rm sh}=T(\lambda),\qquad
x_{\rm sh}=\xi(T)+\frac{\lambda}{2}T-\frac12\!\int^{\lambda}\!\!T\,\dd\lambda',
\qquad
u_{\rm sh}=\frac{\lambda}{2}+V(T).
\label{eq:triplet}
\end{equation}
Since $\dot\xi=V$, these satisfy
$\dd x_{\rm sh}/\dd\lambda=u_{\rm sh}\,\dd t_{\rm sh}/\dd\lambda$ identically, so the shoreline moves with the fluid.

\subsection{Range of validity}

\begin{figure}
  \centering
  \includegraphics[width=\textwidth]{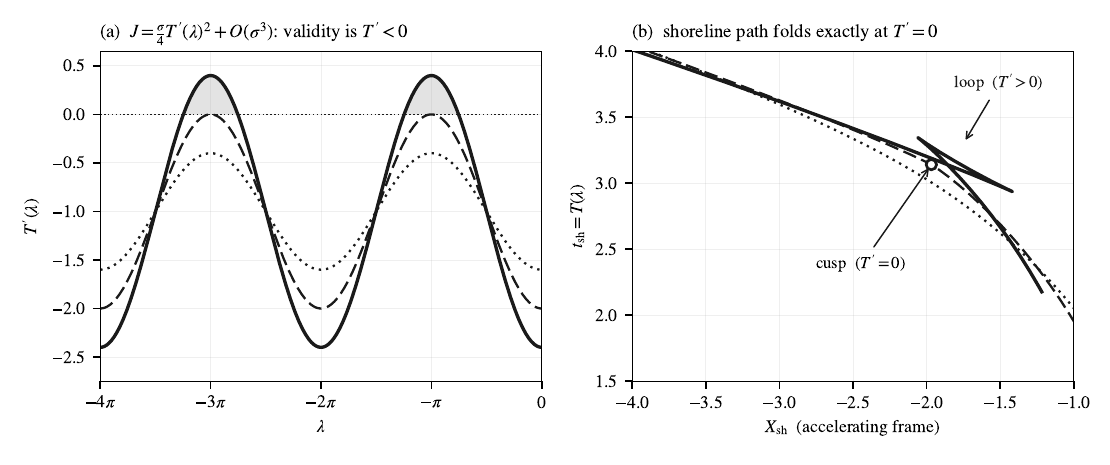}
  \caption{(\textit{a}) $T'(\lambda)$ for $T=-\lambda-A\sin\lambda$ with
  $A=0.6,\,1.0,\,1.4$. Equation \eqref{eq:jac} fixes where the map degenerates and
  \eqref{eq:catastrophe} fixes which side is physical: the undisturbed beach sits on $T'<0$.
  (\textit{b}) The corresponding shoreline paths, drawn in the accelerating
  frame, where the drift of the frame leaves the local structure visible. A cusp
  appears where $T'=0$, and the path folds where $T'>0$.}
  \label{fig:hodograph}
\end{figure}

The Jacobian of the map is $J=c\,(t_\lambda^{2}-t_\sigma^{2})$, and near the shoreline \eqref{eq:texp} reduces it to
\begin{equation}
J=\frac{\sigma}{4}\Bigl[T'(\lambda)^{2}+O(\sigma^{2})\Bigr].
\label{eq:jac}
\end{equation}
At the shoreline the map is one-to-one while $T'$ keeps one sign, and it
degenerates at
\begin{equation}
T'(\lambda)=0 .
\label{eq:valid}
\end{equation}
Figure~\ref{fig:hodograph} is drawn at the shoreline for a sample $T(\lambda)$.
Panel (\textit{a}) plots $T'$ against $\lambda$ at three amplitudes and marks
where it changes sign, and panel (\textit{b}) the shoreline paths that follow,
which develop a cusp at \eqref{eq:valid} and fold beyond it.

This degeneracy is a gradient catastrophe. As $T'$ approaches zero, the
free-surface slope becomes unbounded, and the hodograph map records but does not
cause this behaviour.
Close to the shoreline $\sigma_X=-t_\lambda/J$, so \eqref{eq:texp} and
\eqref{eq:jac} give
\begin{equation}
d_X=\frac{\sigma}{8g}\sigma_X\;\longrightarrow\;-\frac{1}{2gT'} ,
\label{eq:catastrophe}
\end{equation}
which diverges as $T'\to0$ independently of $\sigma$. The undisturbed beach of
slope $\tan\beta$ fixes the sign: there $u=0$ and $\lambda=-2gt\tan\beta$, so
$T'=-1/(2g\tan\beta)$ and \eqref{eq:catastrophe} returns $d_X=\tan\beta$. The
physical branch is $T'<0$. Equation
\eqref{eq:valid} therefore bounds the range of validity and locates breaking at
the shoreline.

Because \eqref{eq:jac} is an expansion about $\sigma=0$, condition
\eqref{eq:valid} is necessary but not sufficient for a one-to-one map.
$J$ can vanish first in the interior; then the wave breaks seaward of the
shoreline. Waves break in this way over a flat bed as well, and the steepening
is governed by the simple-wave argument of \S\,\ref{sec:gen}.

\subsection{The classical limit}

Fix the slope at $\rho\equiv\tan\beta$. Then $V=mt$ and $\xi=\tfrac12mt^{2}$
with $m=g\tan\beta$, and \eqref{eq:triplet} reduces to the shoreline relation of
\citet{CarrierGreenspan1958} (see Appendix~\ref{app:hodo}). The run-up law of
\citet{Synolakis1987} can then be recovered without modification, and the near-shoreline
recursion of \citet{PritchardDickinson2007} carries over with the coefficient
$4n(n+1)$ in place of $4n^{2}$. \citet{ChanLiu2012} carry the forcing term
$g\rho$ of \eqref{eq:char} as $g\,d_x$ in their treatment of run-up on a plane
beach.

\subsection{A first-order law for a tilting beach}\label{sec:law}

The classification also covers a beach that tilts while the wave moves, and
the run-up law for that case follows from \eqref{eq:triplet}. Since the hodograph
solution does not carry the trace of $\rho$, the bottom motion may be perturbed with
$T(\lambda)$ held fixed, that is, $\rho(t)=\tan\beta\,[1+\varepsilon\gamma(t)]$ with $\Gamma(t)=\int_0^{t}\gamma$. The
shoreline position then acquires the increment
$\varepsilon m\!\int_0^{T}\!\Gamma\,\dd t'$, with $m=g\tan\beta$, and the envelope
theorem removes the $O(\varepsilon)$ shift of the maximising $\lambda$.

\begin{theorem}[First-order run-up law]\label{thm:law}
With $T_{\rm m}$ the shoreline time at which the run-up is greatest, the pivot at
the initial shoreline so that $\mathcal{B}\equiv0$, and $R$ the elevation of the waterline
above the initial still-water level,
\begin{equation}
R_{\max}=R^{(0)}_{\max}\bigl[1+\varepsilon\gamma(T_{\rm m})\bigr]
-\varepsilon\,m\tan\beta\!\int_0^{T_{\rm m}}\!\Gamma(t')\,\dd t'+O(\varepsilon^{2}),
\label{eq:law}
\end{equation}
where $R^{(0)}_{\max}$ is the run-up on the fixed beach of slope $\tan\beta$.
\end{theorem}

\noindent The first term gives the quasi-static effect of the instantaneous slope
at maximum run-up. The second shifts the entire shoreline by the accumulated
displacement of the accelerating frame and therefore retains the history of the
bottom motion. Fixing the pivot sets a datum for $R$ and not for the flow, which
theorem~\ref{thm:pivot} leaves untouched. Equation \eqref{eq:law} is written with the hodograph datum
$T(\lambda)$ held fixed. Fixing instead the incident wave in physical space
makes $T$ itself depend on $\varepsilon$ and adds further terms. We note here
that the experimental data in \S\,\ref{sec:comp} were obtained with the fixed
bed, which cannot be used to test \eqref{eq:law}, and the test is left as future
work.

\section{Wave generation}\label{sec:gen}

\subsection{The residual problem}

\begin{figure}
  \centering
  \includegraphics[width=0.72\textwidth]{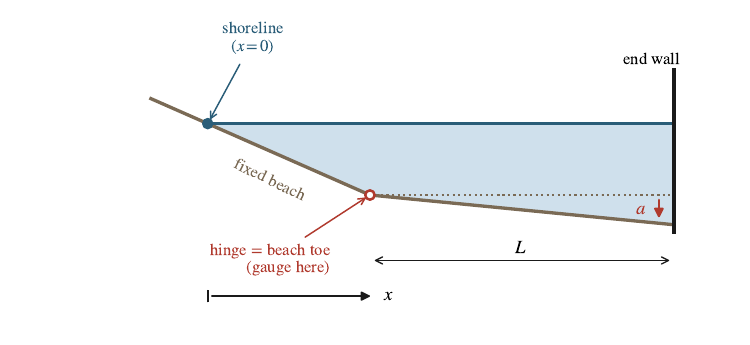}
  \caption{The bottom-tilting wave maker. A flat plate of length $L$ is hinged
  at the toe of a fixed beach, with an end wall at its offshore end and the gauge at
  the hinge; the dotted line is the plate at rest and the solid line the plate
  displaced by $a$. The origin is the initial shoreline and $x$ increases offshore; the drawing is not to scale. With the beach slope set to zero (\S\,\ref{sec:flat}) the bed runs level from the hinge to the far end of the tank and no shoreline exists. The exact solution of \S\,\ref{sec:reduce} applies over the
  plate, and the beach shoreward of the hinge is the classical constant-slope member
  of the same family.}
  \label{fig:apparatus}
\end{figure}

Consider a plate of length $L$ hinged at its shoreward end, with an impermeable end
wall a distance $L$ offshore of the hinge and a still depth $h_0$ at the hinge,
as in figure~\ref{fig:apparatus}.
This is the bottom-tilting wave maker of \citet{LuParkCho2017a, LuParkCho2017b, LuParkCho2020}.

The tank as a whole does not satisfy \eqref{eq:family}. The bottom gradient is
$\rho(t)$ over the moving plate but equals the fixed beach slope shoreward of the
hinge. The gradient $h_x$ therefore varies with $x$. Although theorem~\ref{thm:class} does not apply to the composite,
it can be applied to each region separately. The two parts are
joined at the hinge by continuity of the free surface and of the discharge $d\,u$, and the junction is where the generated wave is measured. The exact solution of
\S\,\ref{sec:reduce} therefore describes the generation region, where the
transformation \eqref{eq:frame} absorbs the uniform forcing into the accelerating
frame. The boundary is then the only inhomogeneity in the residual problem.

\begin{theorem}[Residual boundary problem]\label{thm:piston}
In the accelerating frame \eqref{eq:frame} both boundaries of the tilting region
translate with the same velocity $-V(t)$, so the region moves as a rigid body
and the wall becomes a material boundary. Once \eqref{eq:unforced}
has removed the forcing, the residual is a constant-depth problem, started from
rest and driven by a single moving boundary of velocity $-V(t)$.
\end{theorem}

\noindent Theorem~\ref{thm:piston} turns
the tilting bottom into a problem of constant depth driven at one boundary, and
in the following subsections the closed forms are built on the reduced problem.
Theorem~\ref{thm:pivot} and theorem~\ref{thm:piston} together describe the
bottom-tilting wave maker. The pivot is free to sit at the toe of the beach, and
the plate acts as a piston.

\subsection{The hinge response as a delayed second difference}\label{sec:hinge}

The linear response of this apparatus was obtained by \citet{LuParkCho2017b} as
a Fourier--Laplace double integral with dispersion retained, and
Appendix~\ref{app:disp} recovers it in the present notation. Below, the
non-dispersive limit at the hinge is evaluated in closed form.

The gauge stands at the hinge, and time is measured in transits of the plate,
$\tau=t\sqrt{gh_0}/L$. Let $\mathcal{D}(\tau)$ be the normalised plate
displacement and $G$ the impulse response of the linearised problem at the
hinge. The plate forces the fluid over its whole length, so the hinge responds
from $\tau=0$. Consider a plate section at distance $X$ from the hinge, in units
of $L$. Its signal reaches the gauge directly at $\tau=X$ and, after reflection
from the wall, at $\tau=2-X$. The plate displacement is proportional
to $X$. The two families of arrivals therefore contribute $\tau/2$
over $(0,1)$ and $(2-\tau)/2$ over $(1,2)$, so that
$G'(\tau)=\tfrac12\bigl(1-|\tau-1|\bigr)$ on $(0,2)$, whence
\begin{equation}
G(\tau)=
\begin{cases}
\tfrac14\tau^{2}, & 0\le\tau\le1,\\[4pt]
\tfrac12-\tfrac14(2-\tau)^{2}, & 1\le\tau\le2,\\[4pt]
\tfrac12, & \tau\ge2,
\end{cases}
\label{eq:G}
\end{equation}
and $G(\tau)+G(2-\tau)=\tfrac12$.

\begin{lemmax}[Delayed second difference]\label{lem:delay}
Consider $\mathcal{D}''$ as a distribution on the whole line, so that a start at
constant velocity carries $\mathcal{D}'(0^{+})\delta(\tau)$. Then
$\mathcal{D}''\!*G=\mathcal{D}*G''$, and
$G''=\tfrac12[\Theta(\tau)-2\Theta(\tau-1)+\Theta(\tau-2)]$ with
$\Theta$ the Heaviside step, so the normalised hinge signal is
\begin{equation}
Q(\tau)=\tfrac12\bigl[\mathcal{S}(\tau)-2\mathcal{S}(\tau-1)+\mathcal{S}(\tau-2)\bigr],\qquad
\mathcal{S}(\tau)=\int_0^{\tau}\!\mathcal{D} .
\label{eq:delay}
\end{equation}
\end{lemmax}

\noindent The lag is the transit time of the plate.

\begin{theorem}[Area]\label{thm:area}
If the support of $\mathcal{D}$ is shorter than the transit time, then
$Q=\tfrac12\mathcal{A}\,[\Theta(\tau)-2\Theta(\tau-1)+\Theta(\tau-2)]$ with
$\mathcal{A}=\int_0^{\infty}\mathcal{D}\,\dd\tau$, the crest-to-trough interval taken
between corresponding points of the two extrema is exactly unity, and
\begin{equation}
F^{(0)}=b^{*}\mathcal{A}.
\label{eq:area}
\end{equation}
\end{theorem}

\noindent The plate motion is set by two ratios. One, $a^{*}=a/h_0$, measures the
vertical throw of the plate against the still depth. The other,
$b^{*}=b\sqrt{gh_0}/L$, measures the duration of the plate motion against the
time a long wave takes to cross the plate.

Two plate motions are used. A plate that starts lowered and is raised over a
time $b$ to the horizontal, where it is held, makes a leading-elevation wave. A
plate lowered over a time $b$ and then raised over a further $b$ returns to the
horizontal and makes a leading-depression wave. The
steepness is the mean slope of the front of the hinge record
\citep{LuParkCho2020}. In the units of this section it is
$\kappa^{*}=\zeta_{\rm cr}/\tau_{\rm cr}$ for a leading-elevation wave, measured from the
arrival of the front, and $\kappa^{*}=(\zeta_{\rm cr}-\zeta_{\rm tr})/(\tau_{\rm cr}-\tau_{\rm tr})$ for a
leading-depression wave, the free surface being scaled by $h_0$ and time by $L/\sqrt{gh_0}$. In dimensional variables $\kappa=\alpha c_0\kappa^{*}$ with $c_0=\sqrt{gh_0}$ and
$\alpha=h_0/L$ the relative depth,
so $\kappa$ has the dimensions of velocity and $\kappa/c_0$ is the
dimensionless steepness. In these terms $F^{(0)}=\kappa^{*}b^{*}/a^{*}$ is the
small-amplitude limit. If the plate motion lasts less than one transit time, the
response depends only on the area under the displacement history, not on its shape. A drive that cannot
start or stop instantaneously therefore leaves $F^{(0)}$ unchanged, because any
linear tracking dynamics with a steady gain of unity preserves $\int\mathcal{D}$.

\subsection{Closed form for the linear response}

For a plate raised over a time $b$ and then held, $\mathcal{D}$ is a ramp, the
crest of $Q$ falls at $\tau=1+b^{*}/2$, and
\begin{equation}
F^{(0)}_{\rm LE}(b^{*})=\frac{b^{*}}{2}\,
\frac{1-b^{*}/4}{1+b^{*}/2},\qquad \Delta\tau^{(0)}=1+\frac{b^{*}}{2} .
\label{eq:F0LE}
\end{equation}
The crest of $Q$ is an isolated maximum at every $b^{*}$. The steepness of a
leading-elevation wave is referred to the undisturbed front, so $\Delta\tau^{(0)}$
is the time from the arrival of the front to the crest.

For a plate lowered over a time $b$ and raised over a further $b$,
$\mathcal{D}$ is a triangle of half-width $b^{*}$, and the antisymmetry
$Q(2+2b^{*}-\tau)=-Q(\tau)$ which follows from $G(\tau)+G(2-\tau)=\tfrac12$ gives
two regimes,
\begin{equation}
F^{(0)}_{\rm LD}(b^{*})=
\begin{cases}
(b^{*})^{2}, & b^{*}\le\tfrac12,\\[4pt]
\dfrac{3-(2-b^{*})^{2}}{2+2b^{*}}, & \tfrac12\le b^{*}\le2 .
\end{cases}
\label{eq:F0}
\end{equation}
Below $b^{*}=\tfrac12$ the extrema of $Q$ are exact plateaux, each of width
$1-2b^{*}$, and their centres are one transit time apart. Above it they are
isolated points and the separation opens to $(2+2b^{*})/3$. The plateaux carry a
consequence for measurement. An extremum search may select any point on either
plateau. The measured interval can therefore range from $2b^{*}$ to $2-2b^{*}$,
although the theoretical interval between the plateau centres is unity.

Every leading-elevation entry below uses \eqref{eq:F0LE} and every
leading-depression entry uses \eqref{eq:F0}, and $F^{(0)}$ without a subscript
stands for whichever of the two applies. Dispersion changes $F^{(0)}$ by at most
$4.3\%$ over the beach set and by $5.5\%$ over the
flat-bottom set, whose relative depth is twice as large (Appendix~\ref{app:disp}).

\subsection{Nonlinear correction and the reference point}\label{sec:refpt}

By theorem~\ref{thm:piston} the residual problem has constant depth and starts
from rest. Every $C_-$ characteristic in the disturbed region traces back to
undisturbed fluid ahead of the front, so the invariant $u-2c$ keeps its
quiescent value $-2c_0$ throughout. The motion is a simple wave, and
$c=c_0+u/2$. \citet{MadsenHansen2012} reach a one-family description of a moving
bottom by integrating the backward family out. In the present problem the frame of
theorem~\ref{thm:piston} has already removed the bottom, so the backward family
stays quiescent and the reduction is exact.
In a simple wave each characteristic preserves the surface elevation. The
crest-to-trough height therefore remains unchanged until breaking, while the
propagation speed is $u+c=c_0+\tfrac32u$. The two points
between which the steepness is measured travel at different speeds,
and the interval between them shortens. Keeping the compression exact in the velocity gives
\begin{equation}
\Delta\tau=\Delta\tau^{(0)}+\Bigl[\bigl(1+\tfrac32 u^{*}_{\rm cr}\bigr)^{-1}
-\bigl(1+\tfrac32 u^{*}_{\rm tr}\bigr)^{-1}\Bigr],\qquad u^{*}=u/c_0 ,
\label{eq:dtau}
\end{equation}
in which $u^{*}_{\rm cr}$ and $u^{*}_{\rm tr}$ are the normalised velocities at the crest and
at the trough of the leading wave, and $H$ denotes its crest-to-trough height.

\begin{theorem}[Reference-point asymmetry]\label{thm:refpt}
With $\kappa^{*(0)}=a^{*}F^{(0)}/b^{*}$ the linear steepness and $H=\kappa^{*(0)}\Delta\tau^{(0)}$,
\begin{equation}
\kappa^{*}_{\rm LE}=\frac{\kappa^{*(0)}}{1-\dfrac{\tfrac32\kappa^{*(0)}}{1+\tfrac32H}},
\qquad
\kappa^{*}_{\rm LD}=\frac{\kappa^{*(0)}}{1-\dfrac{\tfrac32\kappa^{*(0)}}{1-\tfrac{9}{16}H^{2}}} .
\label{eq:exact}
\end{equation}
\end{theorem}

\noindent The two cases use different reference points. For a leading-elevation
wave, the measurement begins at the undisturbed front, where the fluid remains at
rest. A leading-depression wave is measured
between two disturbed points whose fluid velocities have opposite signs. The
factors in the two denominators therefore act in opposite directions, and the
compression is weaker for the leading-elevation wave. Both expressions reduce
to $\kappa^{*}=\kappa^{*(0)}/(1-\tfrac32\kappa^{*(0)})$ as $H\to0$, which in the variable $F=\kappa b/a$
reads
\begin{equation}
F=\frac{F^{(0)}(b^{*})}{1-\tfrac32\kappa^{*(0)}} .
\label{eq:first}
\end{equation}
We reiterate here that no constant in \eqref{eq:F0}--\eqref{eq:first} is fitted.

Equation \eqref{eq:exact} retains one approximation. It keeps all orders of
$u^{*}$ in the compression factor but evaluates $u^{*}$ from the linear profile.
The simple-wave relation reads $u^{*}=2(\sqrt{1+\zeta^{*}}-1)=\zeta^{*}+O((\zeta^{*})^{2})$ with $\zeta^{*}=\zeta/h_0$,
and the quadratic term is dropped when $u^{*}_{\rm cr}$ and $u^{*}_{\rm tr}$ are taken as
$H$ and zero for the leading-elevation wave and as $\pm H/2$ for the
leading-depression wave. Equation \eqref{eq:exact} is therefore a resummation of the first-order result rather than an exact expression. We keep the linear
velocity throughout (see \S\,\ref{sec:steep} for the effect of restoring the
quadratic term).

\section{Comparison}\label{sec:comp}

\subsection{The two experiments}\label{sec:data}

We present two sets of experiments which were run on the wave maker of figure~\ref{fig:apparatus}, whose beach slope is adjustable. In the first experiment
the slope was set to zero, so the bed ran level throughout and the
gauge at the hinge recorded the generated wave. In the
second a plane beach filled the shoreward half of the tank, with the plate hinged at
its toe, and the record covers the wave at the hinge together with the run-up on
the beach \citep{LuParkCho2020}. Table~\ref{tab:cond}
summarises experimental cases and conditions.

The two sets have different roles below. The flat-bottom experiment determines
the boundary condition because the hinge record is the incident wave
(\S\,\ref{sec:flat}). The beach experiment then tests how this wave steepens
(\S\,\ref{sec:steep}) and runs up the beach (\S\,\ref{sec:runupcomp}).

We note here that three quantities in these records are of different kinds. The
steepness $\kappa$ of the beach set is a numerical solution of
the nonlinear shallow-water equations, reported without a numerical tolerance,
so we quote relative differences throughout. The free-surface record at the hinge
is a gauge measurement of stated accuracy $\pm0.5$\,mm at $50$\,Hz, which is
$\pm0.017$ when scaled on the still depth at $\alpha=0.03$; \citet{LuParkCho2020}
attribute part of the scatter of the small-amplitude cases to the accuracy of the wave gauge. The run-up is read from video against a $2$\,cm grid marked on the beach. For comparison, our own uncertainty band, from
dispersion and from the far-end reflection, is $5.4\%$
(Appendix~\ref{app:basis}).

\subsection{The generated wave over a flat bottom}\label{sec:flat}

The tank is $2$\,m long, $0.11$\,m wide and $0.2$\,m deep, with bed and walls of
$10$\,mm acrylic; the width makes the motion two-dimensional. A fixed bed fills
one half and the tilting plate of length $L=1$\,m the other, and the two meet at
the hinge in the middle, so
$\ell=2$ in the notation of Appendix~\ref{app:basis}. For this set the fixed half
was left level, so the bed ran level from the hinge to the end of the tank a
distance $L$ away. An
electrical servo motor drove the plate down and back to the horizontal at constant
speed, which generates a leading-depression wave, and \eqref{eq:F0} is
applied. The free surface at the hinge was recorded at $50$\,Hz for
$10$\,s to an accuracy of $\pm0.5$\,mm, and the propagation was filmed from the
side, normal to the tank, at $30$ frames per second and $640\times480$ pixels.
Breaking was identified by two signs, bubbles forming near the crest and a
subsequent sharp decrease in the recorded wave height. Each parameter set was
tested twice to confirm the classification.
Sixty-one independent cases were obtained, over $\alpha=0.04$--$0.08$, $a^{*}=0.38$--$1.25$ and
$b^{*}=0.31$--$1.77$. The
range reaches deeper water, larger amplitude and longer motions than the beach
set (table~\ref{tab:cond}). Nine of the cases were seen to break during propagation. The
crest-to-trough height $H$ and the steepness $\kappa^{*}$ of the leading wave at
the hinge, for all $61$ cases, are reported here for the first time.

\begin{table}
  \centering
  \begin{tabular}{lccc}
  \toprule
   & flat bottom & \multicolumn{2}{c}{plane beach (run-up)}\\
   & (no run-up) & leading-elevation & leading-depression\\
  \midrule
  still depth $h_0$ (m) & $0.040$--$0.080$
    & \multicolumn{2}{c}{$0.020,\ 0.025,\ 0.030$}\\
  relative depth $\alpha=h_0/L$ & $0.040$--$0.080$
    & \multicolumn{2}{c}{$0.020$--$0.030$}\\
  amplitude $a^{*}=a/h_0$ & $0.38$--$1.25$ & $0.17$--$1.00$ & $0.17$--$1.00$\\
  duration $b^{*}=b\sqrt{gh_0}/L$ & $0.31$--$1.77$ & $0.22$--$1.09$
    & $0.22$--$1.09$\\
  beach slope $\tan\beta$ & $0$ & \multicolumn{2}{c}{$1/15,\ 1/20,\ 1/25$}\\
  independent cases & $61$ & $48$ & $52$\\
  \bottomrule
  \end{tabular}
  \caption{Experimental conditions. The plate length is $L=1$\,m and the tank
  length $2$\,m throughout; the three beach slopes are $\beta=3.81^{\circ}$, $2.86^{\circ}$ and $2.29^{\circ}$. The first column is the flat-bottom set reported here
  for the first time; the other two are the data of \citet{LuParkCho2020},
  counted once for each group of cases that differ only in the beach slope.}
  \label{tab:cond}
\end{table}

The set also allows a test that does not require $(a^{*},b^{*})$ to be assigned
case by case. By \S\,\ref{sec:gen} the height is insensitive to the nonlinearity
while the interval is compressed, so the first-order relation \eqref{eq:first}
inverts to
\begin{equation}
\Delta\tau_{\rm lin}=\frac{H}{\kappa^{*}}\Bigl(1+\tfrac32\kappa^{*}\Bigr)=\Delta\tau^{(0)} .
\label{eq:invert}
\end{equation}
Because $H$ cancels in $H/\kappa^{*}$, this ratio gives the crest-to-trough
interval directly. Gauge accuracy affects the ordinate only through the
correction $\tfrac32\kappa^{*}$. The ordinate rests instead on the timing of
the two extrema on a record sampled at $50$\,Hz, and one sample spans $0.013$ to
$0.018$ of a transit time.
Over the range of $b^{*}$, \eqref{eq:F0} admits
$1.00\le\Delta\tau_{\rm lin}\le1.85$. The lower bound is the plate transit time,
which the theory gives for every $b^{*}\le\tfrac12$, and the upper is $(2+2b^{*})/3$ at $b^{*}=1.77$. The
plateaux of \S\,\ref{sec:hinge} loosen that lower edge for the shorter motions,
since an extremum search on a case with $b^{*}<\tfrac12$ can report an interval
as short as $2b^{*}$. The set reaches $b^{*}=0.31$, so $0.62$ is the shortest
interval any case in the set can return regardless of the duration.

\begin{figure}
  \centering
  \includegraphics[width=\textwidth]{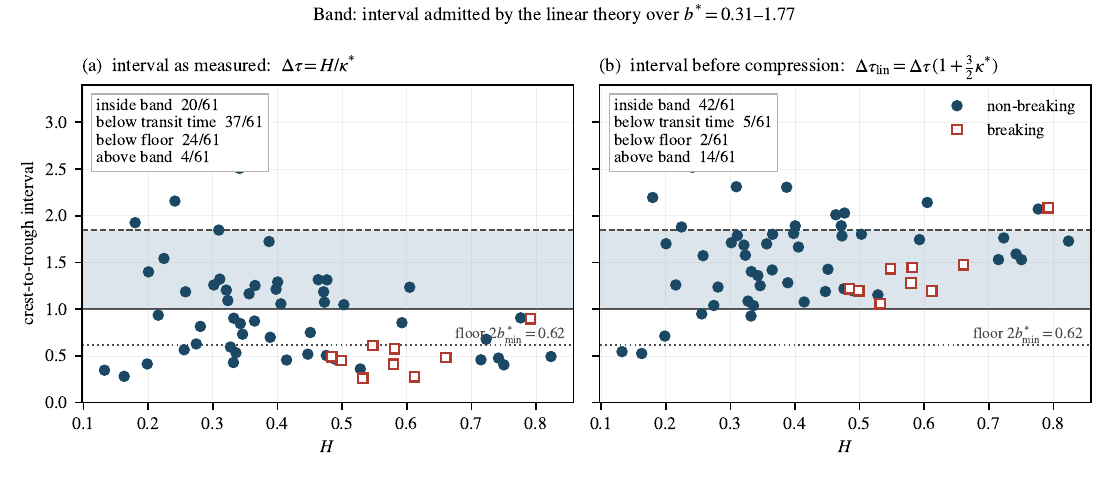}
  \caption{The flat-bottom set. The solid line is the plate transit time, which
  the linear theory gives for every motion with $b^{*}\le\tfrac12$, and the
  dotted line is the floor $2b^{*}_{\min}=0.62$ set by the plateaux of
  \S\,\ref{sec:hinge}, which binds every case whatever its duration.
  (\textit{a}) Without the compression the inferred interval falls below the
  transit time for $37$ of the $61$ cases and below the floor for $24$ of them.
  (\textit{b}) Undoing the compression through \eqref{eq:invert} lifts the
  population into the admitted band, leaving $2$ below the floor and $14$ above
  the upper edge. Open squares
  mark the cases seen to break, which for this set is an observation rather than
  a criterion.}
  \label{fig:unpub}
\end{figure}

Figure~\ref{fig:unpub} shows the outcome. Without the compression the median is
$0.85$, and $37$ of the $61$ cases call for a crest-to-trough interval shorter
than the plate transit time. Of those, $24$ fall below the floor $0.62$, which no
case in the set can return under any duration. Correcting for first-order
compression raises the median to $1.53$, with quartiles of $1.22$ and $1.80$. Of
the $61$ cases, $42$ then lie within the predicted band, $2$ below it and $14$
above it. The nine breaking cases sit among the rest. The factor $1+\tfrac32\kappa^{*}$ runs from $1.14$ to
$4.32$ across the set, and the one constant multiplier that puts the most cases
inside the band leaves $26$ of the $61$ there, against the $42$ that the
case-by-case correction places.

It is noted here that equation \eqref{eq:invert} inverts the first-order
relation. Inverting the leading-depression branch of
\eqref{eq:exact} instead adds a factor $(1-\tfrac{9}{16}H^{2})^{-1}$ to the
correction. That raises the median to $1.71$ and leaves $33$ of the $61$ inside
the band, with $23$ above the upper edge. The beach set shows the same over-compression at
large amplitude, which will be discussed in \S\,\ref{sec:steep}.

\subsection{Steepness at the toe of the beach}\label{sec:steep}

The beach set was taken in a tank $2$\,m long, with a moving bottom over half of
it, $L=1$\,m, hinged at the toe of a plane beach that fills the other half.
Three still depths were used, $h_0=0.02$, $0.025$ and $0.030$\,m, and three beach
slopes, $1/15$, $1/20$ and $1/25$. The plate throw and the duration of its motion
cover $a^{*}=0.17$--$1.00$ and $b^{*}=0.22$--$1.09$. The beach slope does not enter the generation problem, so cases that share
$(a^{*},b^{*})$ and differ only in the beach slope repeat one generation
condition. Counting each repeated generation condition only once leaves $48$
independent leading-elevation cases and $52$ leading-depression cases.

\begin{figure}
  \centering
  \includegraphics[width=\textwidth]{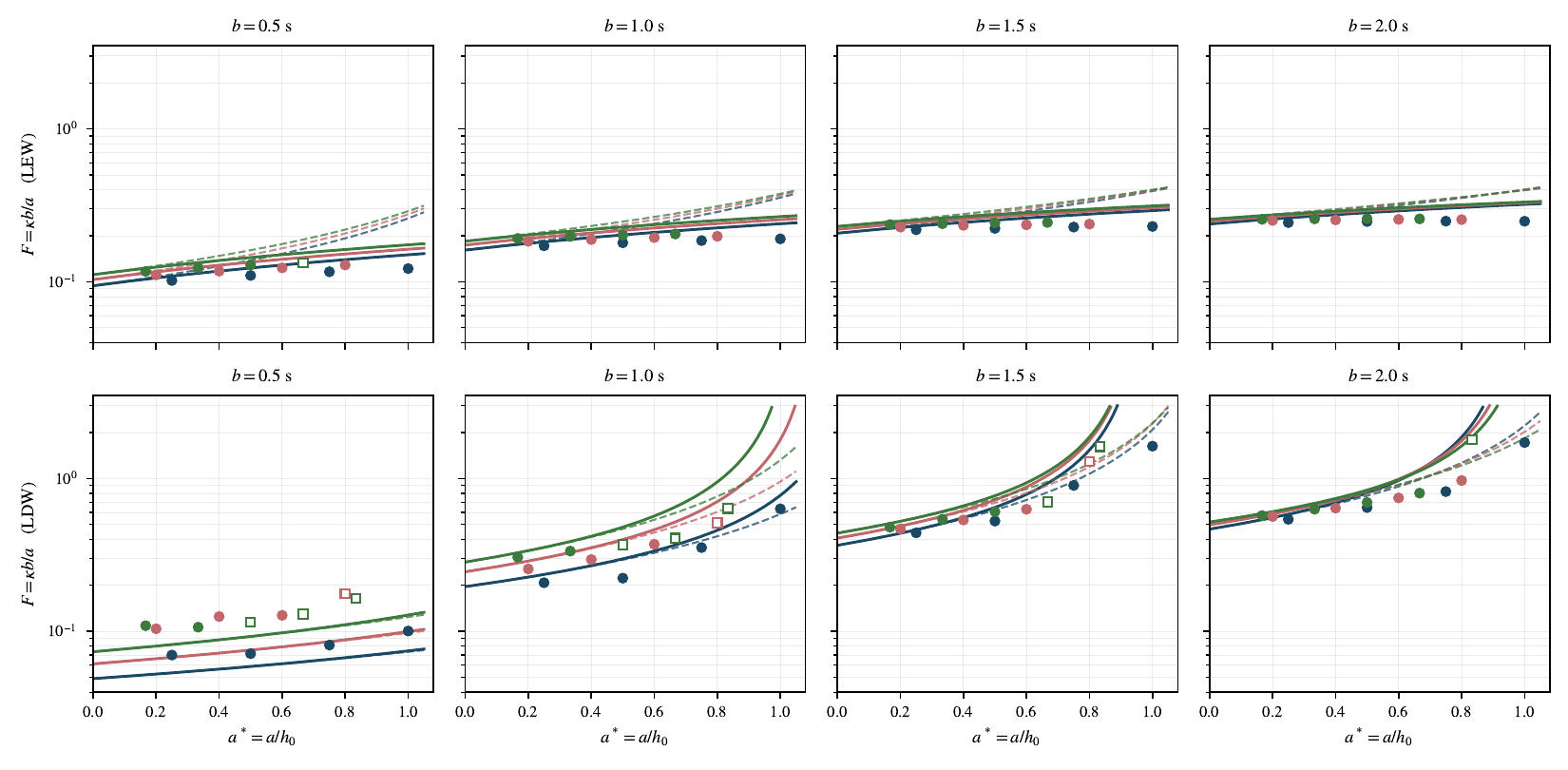}
  \caption{$F=\kappa b/a$ against $a^{*}$, compared with the reference computation of \citet{LuParkCho2020}. There is no
  measurement in this figure. Upper row, leading-elevation waves;
  lower row, leading-depression waves; columns, the commanded duration $b$.
  Filled circles are cases not classified as breaking, and open squares are cases classified as breaking on the beach by the criterion
  of \citet{MadsenSchaffer2010}; the latter are retained because the steepness is
  measured at the hinge, seaward of the beach. Colour denotes the still depth:
  dark blue $h_0=0.020$\,m, red $0.025$\,m, green $0.030$\,m, and the curves
  take the colour of the depth they belong to. Broken lines, first-order compression \eqref{eq:first}; solid lines, the reference-point correction
  \eqref{eq:exact}. The uncertainty band of $5.4\%$ is not drawn here, since the reference is a
  non-dispersive computation in a semi-infinite tank and neither contribution to
  the band applies. The
  ordinate is logarithmic and common to all panels, so equal vertical
  separations correspond to equal ratios. No constant is fitted.}
  \label{fig:F}
\end{figure}

Figure~\ref{fig:F} shows the comparison of the predicted steepness with the
reference computation of \citet{LuParkCho2020}. The relative differences reported below are (prediction $-$ reference)/reference,
so a positive value places the prediction above the reference. Cases that break on the beach are kept here. Breaking is assigned there by the
linear criterion of \citet{MadsenSchaffer2010} as applied by
\citet{LuParkCho2020}, who state that the assignment does not rest on
observation. The steepness is measured at the hinge, seaward of
the beach, so the breaking does not reach it.

Take the leading-elevation wave first. At $a^{*}\le0.4$ the median relative
difference over $20$ cases is $8.7\%$ for the first-order form \eqref{eq:first}
and $6.5\%$ for \eqref{eq:exact}, with an interquartile range of $5.0$--$8.9\%$.
At $0.4<a^{*}\le0.7$ the two forms give $25.8\%$ and $14.7\%$ over $16$ cases.
At $a^{*}>0.7$ they give $53.3\%$ and $21.5\%$ over $12$ cases. The reference-point correction holds over the whole amplitude range. The ratio of reference to prediction follows one gentle decline for the
leading-elevation cases, from $0.96$ at small amplitude to $0.79$ at $a^{*}=1$.
The decline is the same at every commanded duration.

The leading-depression wave behaves differently at large amplitude. At
$a^{*}\le0.4$ the two forms give $11.9\%$ and $13.0\%$ over $20$ cases. At
$a^{*}>0.4$ they give $21.8\%$ and $31.7\%$ over $30$ cases, so \eqref{eq:exact}
makes the difference worse. The simple-wave extrapolation over-compresses there.
Part of the excess belongs to the resummed form itself. The denominator of the
leading-depression branch of \eqref{eq:exact} vanishes at
$a^{*}\simeq0.96$--$1.10$ for $b^{*}\ge\tfrac12$. Two of the $52$ cases, both at
$a^{*}=1.00$, lie beyond the pole and return a negative steepness. The
leading-elevation branch has no pole over the same range. The contrast between
the two waves at large amplitude is therefore in part algebraic. The slope of the reference computation agrees with the coefficient $\tfrac32$ in \eqref{eq:first} to within $12.0\%$ on average over the twelve intervals.
At large amplitudes the reference results show weaker compression, suggesting
that the steepening saturates.

The velocity in \eqref{eq:exact} is read from the linear profile. Restoring the
quadratic term of the simple-wave relation moves the three leading-elevation
medians to $6.2$, $13.6$ and $19.4\%$ and raises the leading-depression
difference at $a^{*}>0.4$ to $36.7\%$.

Grouped by commanded duration at $a^{*}\le0.4$, the median differences are $+6.8$,
$+6.2$, $+5.7$ and $+7.4\%$ for the leading-elevation wave at
$b=0.5,\,1.0,\,1.5,\,2.0$\,s. For the leading-depression wave they are $-27.6$,
$+13.3$, $+5.9$ and $+7.9\%$. A positive offset of $5.7$--$7.4\%$ runs through the leading-elevation set at every
$b$, and one of $5.9$--$13.3\%$ through the leading-depression set for $b\ge1.0$\,s. Its cause is an
open question. The shortest commanded motion of the leading-depression set is
explained by \S\,\ref{sec:hinge}. These cases have $b^{*}=0.22$--$0.27$, the only
range of the experiment below $\tfrac12$, and they are the only responses that
contain plateaux. Because of these flat regions an extremum search may report
an interval as short as $2b^{*}$ in place of the expected unity, which makes
the steepness read from the record larger by as much as a factor $1/(2b^{*})$. The
observed increase, a factor of $1.4$, lies inside that range. The
leading-elevation wave falls in the same range of $b^{*}$ and shows no such
behaviour.

Both datasets agree with the linear prediction. The nonlinear correction holds
for the leading-elevation wave at all tested amplitudes, but only up to
$a^{*}\simeq0.4$ for the leading-depression wave. At larger amplitudes the
compression stops increasing, for reasons that remain unclear.

\subsection{Run-up}\label{sec:runupcomp}

The beach in this tank is fixed, so the run-up tests the classical member of the
family. The comparison also tests the theory over a longer path. The closed
forms of \S\,\ref{sec:gen} supply the incident wave at the toe, and the
plane-beach solution gives the run-up. \citet{Synolakis1987} used the same
approach for the solitary wave.

Let $\zeta_{\rm inc}(t)$ be the incident free surface at the toe. Let
$t_0=x_0/\sqrt{gh_0}$ be half the long-wave travel time over the beach, whose toe
stands at $x_0=h_0/\tan\beta$. The factor of two comes from the shoaling depth,
since $\int_0^{x_0}(gx\tan\beta)^{-1/2}\dd x=2x_0/\sqrt{gh_0}$. The linear
transfer of \citet{KellerKeller1964} across the toe,
combined with the hodograph solution on the beach, gives the shoreline elevation
as
\begin{equation}
\hat R(\omega)=\frac{2\,\hat\zeta_{\rm inc}(\omega)}
{J_0(2\omega t_0)-\mathrm{i}J_1(2\omega t_0)} ,
\label{eq:transfer}
\end{equation}
\noindent in which a hat denotes the Fourier transform in time, $\omega$ is the
angular frequency, and $J_0$ and $J_1$ are the Bessel functions of the first kind.
For $2\omega t_0\gg1$ the transfer reduces to a half-derivative
\citep{Synolakis1987, MadsenSchaffer2010},
\begin{equation}
R(t)=\sqrt{4\pi t_0}\;\partial_t^{1/2}\zeta_{\rm inc}(t-2t_0),\qquad
\partial_t^{1/2}\leftrightarrow(-\mathrm{i}\omega)^{1/2} .
\label{eq:halfderiv}
\end{equation}
The maximum can be found from \eqref{eq:halfderiv}. The shoreline elevation
changes at the rate $-u_{\rm sh}\tan\beta$, so the shoreline velocity $u_{\rm sh}$ vanishes where $R$
is greatest. Therefore, the quadratic term $-u_{\rm sh}^{2}/(2g)$, included by
\citet{MadsenSchaffer2010} alongside \eqref{eq:halfderiv}, does not affect the
maximum run-up $R_{\max}$. The maximum run-up depends linearly on the incoming
wave. As \citet{Synolakis1987} showed, this means that the linear and nonlinear
theories predict the same maximum run-up.

Substituting the hinge signal of \S\,\ref{sec:hinge} gives a closed form. Write
$\Pi(b^{*})=\max_\tau\partial_\tau^{1/2}Q$ for the half-derivative peak of that
signal. The half-derivative weights the front of the signal, so the rising
part of the incident wave sets the run-up. \citet{PujaraEtAl2020} reach the
same part of the wave from the laboratory, finding that the shape of the
acceleration phase governs the swash that follows. Then
\begin{equation}
\frac{R_{\max}}{h_0}=a^{*}\Pi(b^{*})
\Bigl(\frac{4\pi\alpha}{\tan\beta}\Bigr)^{1/2},
\label{eq:runup}
\end{equation}
and since $\kappa^{*}=a^{*}F^{(0)}/b^{*}$ the run-up is proportional to the predicted
steepness,
\begin{equation}
\mathcal{R}\equiv\frac{R_{\max}}{h_0}
\Bigl(\frac{\tan\beta}{4\pi\alpha}\Bigr)^{1/2}=\Lambda(b^{*})\,\kappa^{*},\qquad
\Lambda(b^{*})=\frac{b^{*}\Pi(b^{*})}{F^{(0)}(b^{*})} .
\label{eq:runupline}
\end{equation}
The beach slope affects the result only through $(4\pi\alpha/\tan\beta)^{1/2}$,
which describes the beach's geometric amplification. The duration of the plate
motion affects the result only through $\Lambda$. Over the range of
table~\ref{tab:cond}, $\Lambda$ runs from
$1.25$ to $1.61$ for the leading-elevation wave. For the leading-depression wave it
is not monotone: it falls from $1.58$ to $0.95$ at $b^{*}\simeq0.56$ and returns to
$1.12$. No constant is fitted.

The approximation in \eqref{eq:halfderiv} is not valid for these tests. With
$\omega$ set by the crest-to-trough interval, the parameter $2\omega t_0$ ranges
from $1.3$ to $4.2$ over the beach set, below the value of $4.88$ that
\citet{MadsenSchaffer2010} give for an error under $5\%$. We therefore evaluate
the full equation \eqref{eq:transfer} numerically. Compared with this result,
\eqref{eq:runup} underestimates the leading-elevation wave by $8.5\%$ and
overestimates the leading-depression wave by $8.7\%$. This difference is due to
their different frequency spectra.

\begin{figure}
  \centering
  \includegraphics[width=\textwidth]{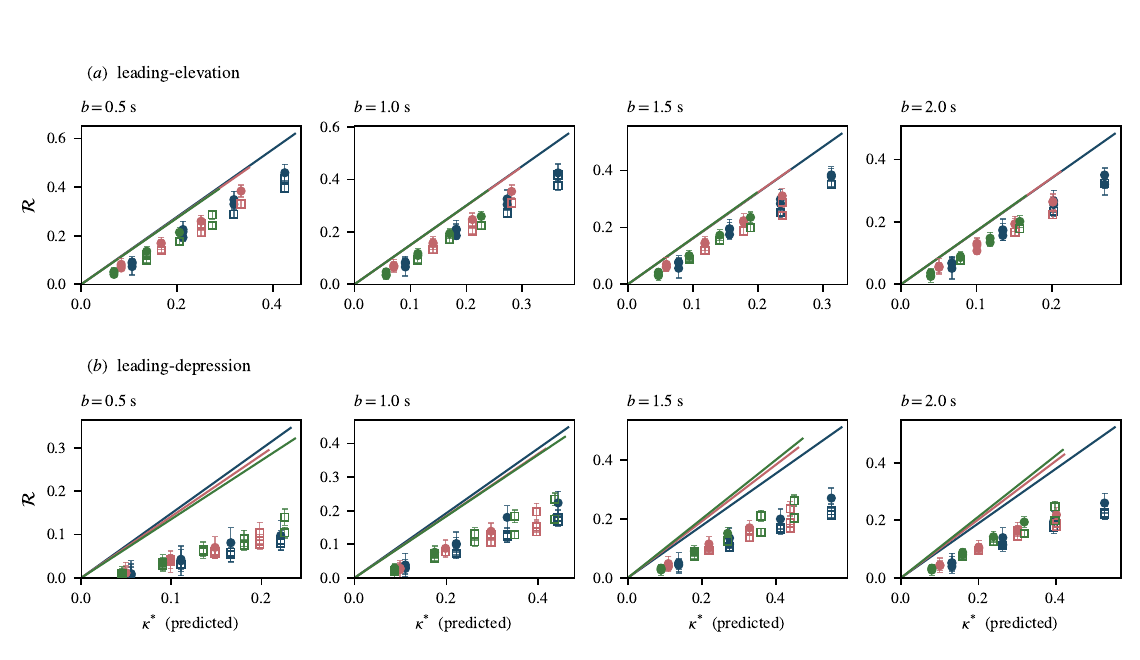}
  \caption{Measured run-up against the predicted steepness. Upper row,
  leading-elevation waves; lower row, leading-depression waves; columns, the
  commanded duration $b$. The ordinate $\mathcal{R}$ is the measured vertical
  run-up with the geometric amplification of the beach divided out, so that
  \eqref{eq:runupline} is a straight line through the origin.
  Filled circles are non-breaking cases and open squares are cases classified as breaking on the beach; colour denotes the still depth, dark blue $h_0=0.020$\,m, red
  $0.025$\,m, green $0.030$\,m. Solid lines are \eqref{eq:runupline}, drawn in the colour
  of the depth they belong to, with $\Lambda$ taken from \eqref{eq:transfer} at
  $\cot\beta=20$ and no constant fitted. Bars carry one grid spacing of the video reading, $2$\,cm along the
  beach face, brought onto the ordinate. Open squares are
  drawn but are left out of the ratios quoted in the text.}
  \label{fig:runup}
\end{figure}

Figure~\ref{fig:runup} compares the measurements against \eqref{eq:runupline}. The
solid lines carry the whole dependence on plate throw, motion duration, still
depth and beach slope. The measurements follow the lines and lie below them.

Two groups of cases are excluded. Beach-slope tests are counted separately
because the slope affects run-up. Of the $144$ leading-elevation records, $10$
lack run-up measurements. Of the remaining $134$, the breaking criterion removes
$37$ and the measurement threshold removes $7$, leaving $90$, distributed as
$41$, $33$ and $16$ over $\cot\beta=15$, $20$ and $25$. Of the $152$
leading-depression records, $8$ lack run-up measurements. Of the remaining
$144$, the breaking criterion removes $75$ and the measurement threshold removes
$33$, leaving $36$, distributed as $25$, $8$ and $3$.
The reading is of the horizontal swash distance, from which the vertical run-up
follows on multiplication by $\tan\beta$. The camera was mounted above the beach
with its image plane parallel to the beach face and the $2$\,cm grid was marked on
that face, so the reading is a distance along the face and needs no parallax
correction; the maximum run-up is the furthest point the front reaches. A minimum vertical run-up of
$2.7$\,mm is used because smaller values have a video-measurement uncertainty of
at least one-third. A horizontal threshold would represent different vertical
heights on different slopes and would exclude more cases on steeper beaches. The
vertical threshold avoids this problem but may still bias the results because it
is applied to the measured outcome. Over what
remains, the ratio of predicted to measured run-up is $1.39$ for the
leading-elevation wave and $1.8$--$2.0$ for the leading-depression wave, with
scatter $13.9\%$ and $19.4\%$. Those two scatters are close to what the reading
alone would give. One grid line spans $2$\,cm on the beach face. The swash distances that survive
the threshold have medians of $7.9$ grid spacings for the leading-elevation
wave and $5.0$ for the leading-depression wave, so a one-spacing uncertainty
amounts to $12.7\%$ and $20.0\%$. The leading-depression
waves have shorter run-up, so the same reading error produces greater relative
scatter. This measurement uncertainty is shown by the error bars in
figure~\ref{fig:runup}, and the flow does not cause it.
The leading-elevation figure holds wherever the
threshold is placed, moving only between $1.34$ and $1.41$ as it runs from zero
to $10$\,mm. The leading-depression figure moves with it, and the range quoted
is what the threshold gives between $2.0$ and $3.5$\,mm; its sample is the
smaller of the two and its run-up the shorter, so the threshold cuts more of it.

The ratio also carries a dependence on the beach slope, which the totals hide
(table~\ref{tab:slope}). The leading-depression column varies by more than the
scatter quoted above, over samples of $25$, $8$ and $3$.

\begin{table}
  \centering
  \begin{tabular}{cccccc}
  \toprule
   & \multicolumn{2}{c}{leading-elevation} & & \multicolumn{2}{c}{leading-depression}\\
  \cmidrule{2-3}\cmidrule{5-6}
  $\cot\beta$ & $n$ & ratio & & $n$ & ratio\\
  \midrule
  $15$ & $41$ & $1.36$ & & $25$ & $1.76$\\
  $20$ & $33$ & $1.37$ & & $8$  & $2.14$\\
  $25$ & $16$ & $1.61$ & & $3$  & $2.67$\\
  \bottomrule
  \end{tabular}
  \caption{Ratio of predicted to measured run-up by beach slope, over the cases
  that pass the breaking criterion and the reading threshold.}
  \label{tab:slope}
\end{table}

The present solution omits friction, and \citet{LuParkCho2020} reported the same
excess in their own inviscid computation, which they removed with a quadratic drag
term. Bottom friction is the candidate for the offset. The present data do not
settle it. For leading-elevation waves the excess decreases with both amplitude
and beach slope. After removing the slope trend, the residual correlates with
$\log a^{*}$ at $r=-0.64$, which accounts for two fifths of its variance.
Controlling for amplitude reduces the slope coefficient from $-0.32$ to
$-0.12$. Three beach slopes cannot separate the two effects,
and the mildest slope of the leading-depression set
holds three records. No exponent in $\tan\beta$ is claimed.

The run-up also bears on the surf-similarity parameter. \citet{LuParkCho2020} organised these data by collapsing the amplification $R_{\max}/\zeta_0$, with $\zeta_0$ the incident
amplitude at the toe, against $\tan\beta\,(\kappa/c_0)^{-1/2}$, the surf-similarity
parameter that \citet{Battjes1974} established for periodic waves. Replacing the
reference $\kappa$ by the present prediction preserves the collapse. The scatter about a common trend is $12.6\%$ against
$12.7\%$ for the leading-elevation cases ($n=97$), and $21.4\%$ against $21.8\%$
for the leading-depression cases at $a^{*}\le0.4$ ($n=45$). Those two counts
follow the breaking criterion alone, without the resolution rule, since the
scatter statistic uses the amplification and not the run-up level. Neither pair is
separated by more than the sampling uncertainty of a scatter estimate. Similar
scatter does not establish equivalence, and it only shows that the two inputs
cannot be distinguished with these sample sizes. Equation \eqref{eq:runupline} is
more predictive than the empirical collapse. The collapse fits two constants to
estimate the amplification, whereas \eqref{eq:runupline} predicts the run-up
directly from the plate motion.

\section{Conclusions}\label{sec:concl}

The Carrier--Greenspan linearisation remains exact for a moving bottom if and
only if the bottom gradient is uniform in space. The admissible family is
$h=\rho(t)x+\mathcal{B}(t)$. Constant slope, the
case of 1958, is one member of the family. For this family a horizontally accelerating
frame recovers the unforced shallow-water equations, so a tilting bottom is
dynamically equivalent to a spatially uniform horizontal body force. One hodograph equation serves
the whole family, and the physical plane follows by quadratures. A varying
$\rho$ rules out an infinite domain, and the pivot is a
choice of datum that leaves the flow unchanged.

A bottom-tilting wave maker realises this narrow family. The linear response at
the hinge is a delayed second difference of the
integrated plate displacement, and the small-amplitude steepness has the closed
form \eqref{eq:F0}. The leading nonlinear correction \eqref{eq:exact} depends on
the point to which the measurement is referred. Over the range set out in
\S\,\ref{sec:comp} the predictions agree with the laboratory data without a
fitted constant. An independent flat-bottom dataset, covering deeper
water and larger amplitudes, supports the proposed compression mechanism.

Because the hinge is the toe of the beach, the same closed forms deliver the
incident wave that the classical plane-beach solution needs. The plate motion
alone therefore determines the predicted run-up. The prediction exceeds the
measured run-up by a median factor of $1.39$ for
leading-elevation waves and $1.8$--$2.0$ for leading-depression waves. Bottom friction is the candidate for the offset, and
the present data do not settle it.

Three questions are left open. First, the mechanism that saturates the compression at
large amplitude remains to be identified. Second, an offset of $5.7$--$13.3\%$ between
the present solutions and the reference computations remains to be explained.
Third, the classification provides a first-order run-up law for a beach that
itself tilts. The law awaits an experiment in which the beach is the moving
element.

\begin{bmhead}[Acknowledgements]
The laboratory experiments were carried out in the Fluids Laboratory at the
University of Dundee, Scotland. The author thanks Heng Lu, who carried out the experiments.
\end{bmhead}

\begin{bmhead}[Funding]
This research was supported by the National Research Foundation of Korea (NRF)
grant funded by the Korea government (MSIT) [RS-2026-25475182]. The author also
acknowledges support from the Institute of Engineering Research at Seoul
National University.
\end{bmhead}

\begin{bmhead}[Declaration of interests]
The author reports no conflict of interest.
\end{bmhead}

\begin{bmhead}[Data availability statement]
The data that support the findings of this study are available in the
supplementary material. The crest-to-trough height $H$, the steepness
$\kappa^{*}$ and the breaking classification are listed there for all $61$
flat-bottom cases. The free-surface records from which those quantities were
taken are no longer retained. The data of \citet{LuParkCho2020} are available
in that paper.
\end{bmhead}

\appendix

\section{Hodograph algebra}\label{app:hodo}

\emph{Reduction to the Riemann variables.} Write $r=u+2c$ and $\tilde r=u-2c$. Along a
$C_+$ characteristic, \eqref{eq:char} fixes $\dd r/\dd t=g h_x$. In contrast,
$\dd\tilde r/\dd t=g h_x+2c\,\tilde r_x$ is unrestricted because the initial data
may assign any value to $\tilde r_x$ at a point where $c>0$. A function $P_+(r,\tilde r,x,t)$ that
stays constant along $C_+$ for every solution therefore needs
$\partial P_+/\partial\tilde r=0$. The Riemann variable is forced rather than assumed.
The argument runs on one family at a time. Imposed on both families at once, the
requirement would leave only $P(x,t)$, and $P_x(u\pm c)+P_t=0$ would then force
$P$ constant.

\emph{Separation in $x$.} With $P_+=P_+(r,x,t)$, constancy along $C_+$ yields
$P_r\,g h_x+P_x(u+c)+P_t=0$. Hold $r$ fixed and vary $c$. Since $u+c=r-c$ and
the three derivatives do not depend on $c$, the coefficient of $c$ forces
$P_x=0$, and $P_r(r,t)\,g h_x(x,t)=-P_t(r,t)$ remains. Non-triviality,
$P_r\neq0$, then forces $h_x$ to be independent of $x$. The same argument on
$C_-$ leads to the same condition.

\emph{Relation to the classical potential.} With
$E_\nu[y]=y_{\sigma\sigma}+\nu\sigma^{-1}y_\sigma-y_{\lambda\lambda}$, if
$E_1[\psi]=0$ then $y=\psi_\sigma/\sigma$ satisfies $E_3[y]=0$. Writing the
regular expansions $\psi=\sum b_n\sigma^{2n}$ and $t=\sum a_n\sigma^{2n}$ gives
$b_{n+1}=b_n''/[4(n+1)^{2}]$ against $a_{n+1}=a_n''/[4(n+1)(n+2)]$. The two are
joined by
\begin{equation}
t=-\frac{\psi_\sigma}{\sigma}-\frac{\lambda}{2m},
\label{eq:dict}
\end{equation}
in which the second term solves $E_3$ on its own. \citet{CarrierGreenspan1958} write
the same relation as $t=\lambda/2-u$.
Setting $\sigma=0$ in \eqref{eq:dict} leaves the
dictionary $T=-\tfrac12\psi_{\lambda\lambda}(\lambda,0)-\lambda/(2m)$, with $m=g\tan\beta$ for a beach of constant slope $\tan\beta$.

\section{Dispersive linear response}\label{app:disp}

Write $s$ for the Laplace variable in time and $k$ for the wavenumber, and let
$\hat h_b(k,s)$ be the transform of the bottom displacement, which factorises
as $a\hat f(k)$ times the transform of $\mathcal{D}(\tau)$. The wall
enters through the even extension used in \S\,\ref{sec:hinge}, so that
$\hat f=L\,\mathrm{sinc}^{2}(kL/2)$, with $\mathrm{sinc}\,x=\sin x/x$ and the transform taken about the wall. The linear response with dispersion retained
is then $\hat\zeta=s^{2}\hat h_b/[\cosh(kh_0)(s^{2}+\omega^{2})]$ with
$\omega^{2}=gk\tanh(kh_0)$, and $F^{(0)}$ follows from the inverted signal at the
hinge by the same crest-to-trough reading as in \S\,\ref{sec:hinge}. Over
$\alpha=0.02$--$0.03$ the dispersive $F^{(0)}_{\rm LD}$ departs from \eqref{eq:F0} by
$-3.8\%$ to $+4.3\%$, and by less than $0.7\%$ for $b^{*}\ge0.66$. Dispersion
mainly changes the crest-to-trough interval, and it changes the amplitude by less
than $1\%$.

The flat-bottom set reaches $\alpha=0.04$--$0.08$, where dispersion is
stronger. Evaluating the same integral over the combinations of $\alpha$ and
$b^{*}$ actually run gives a largest departure of $5.5\%$, at $b^{*}=0.58$, and
$5.4\%$ at the shortest motion $b^{*}=0.31$, falling to $2.9\%$ or less for
$b^{*}\ge0.66$. The effect is therefore concentrated in the shorter plate motions.

\section{Basis of comparison}\label{app:basis}

The far-end reflection reaches the hinge at $\tau=2(\ell-1)$, where $\ell$ is
the tank length in units of $L$; for $\ell=2$ this coincides with the transit
structure of \eqref{eq:G}. Evaluating the leading wave with the two-wall image series changes $F^{(0)}$ by less
than $0.2\%$ for $b^{*}\le\tfrac12$ and by at most $3.3\%$ over
$\tfrac12<b^{*}\le1.09$, which covers the beach set. The flat-bottom set reaches
$b^{*}=1.77$, where the trough of the leading wave falls at
$2(2+2b^{*})/3=3.7$ transit times and the reflection has arrived at $\tau=2$.
The bound is not established there, and the band of \S\,\ref{sec:data} carries
that limitation at its upper edge. The dispersion and reflection errors vary in
opposite directions with $b^{*}$, so their maximum values do not occur
together. Below $b^{*}=\tfrac12$ the reflection contributes less
than $0.2\%$ and the band is set by dispersion at $4.3\%$; above $b^{*}=0.66$
dispersion contributes less than $0.7\%$ and the band is set by the reflection
at $3.3\%$. Combining them in quadrature gives
$5.4\%$, the figure used in \S\,\ref{sec:comp}; adding them yields $7.6\%$.
Neither value is reached at any $b^{*}$. The data are consistent with a reflecting far end. An open
far boundary would put the leading-depression response several times above the
observed values.

\bibliographystyle{jfm}
\bibliography{manuscript}

\end{document}